\documentclass[pra,aps,twocolumn]{revtex4}
\usepackage{amssymb}
\usepackage{latexsym}

\newcommand{\newsec}{\setcounter{equation}{0}\section}

\newcommand{\R}{{\mathbb R}}

\def\be{\begin{equation}}
\def\ee{\end{equation}}
\def\bea{\begin{eqnarray}}
\def\eea{\end{eqnarray}}

\def\d{{\,\rm d}}

\def\k{{\bf k}}

\def\q{{\bf q}}

\def\r{{\bf r}}
\def\t{{\bf t}}
\def\p{{\bf p}}

\def\x{{\bf x}}
\def\y{{\bf y}}

\def\bold0{{\bf 0}}

\def\veps{\varepsilon}
\def\h2m{\frac{\hbar^2}{2m}}
\def\p0{{P_{\beta H^0_N}}}

\def\tilda{{\widetilde{a}}}

\def\Hnot0{{H_{\neg 0}}}

\begin{document}

\title{\vspace{1cm} \large\bf Variational wave functions for homogenous Bose systems}
\author{Andr\'as S\"ut\H o$^1$ and P\'eter Sz\'epfalusy$^{1,2}$\\
$^1$Research Institute for Solid State Physics and Optics, Hungarian Academy of Sciences,\\ P. O. Box 49, H-1525 Budapest, Hungary\\
$^2$Department of Physics of Complex Systems, E\"otv\"os University, H-1117 Budapest, Hungary
}
\thispagestyle{empty}
\begin{abstract}
\noindent
We study variational wave functions of the product form, factorizing according to the wave vectors $\k$, for the ground state of a system of bosons interacting via positive pair interactions with a positive Fourier transform. Our trial functions are members of different orthonormal bases in Fock space. Each basis contains a quasiparticle vacuum state and states with an arbitrary finite number of quasiparticles. One of the bases is that of Valatin and Butler (VB), introduced fifty years ago and parametrized by an infinite set of variables determining Bogoliubov's canonical transformation for each $\k$. In another case, inspired by Nozi\`eres and Saint James the canonical transformation for $\k=0$ is replaced by a shift in the creation/annihilation operators. For the VB basis we prove that the lowest energy is obtained in a state with $\sim\sqrt{\rm volume}$ quasiparticles in the zero mode. The number of $\k=0$ physical particles is of the order of the volume and its fluctuation is anomalously large, resulting in an excess energy. The same fluctuation is normal in the second type of optimized bases, the minimum energy is smaller and is attained in a vacuum state. Associated quasiparticle theories and questions about the gap in their spectrum are also discussed.

\vspace{2mm} \noindent PACS: 03.75.Hh, 05.30.Jp

\vspace{2mm}
\noindent

\end{abstract}
\maketitle

\newsec{Introduction}

Since now more than half a century the understanding of Bose-Einstein condensation (BEC) in interacting Bose gases has been representing a major challenge for theoretical physics. In spite of a considerable progress, the decisive step in rigorous theory | simply, the proof that BEC indeed occurs | still could not be made. Approximate or effective theories that take BEC for granted have also been struggling with some emblematic problems as, for example, the question of the gap appearing almost inevitably in them. The search for gapless approximate theories has a long history that will not be evoked here; instead, we refer the reader to the review of Zagrebnov and Bru \cite{ZB} and to the introduction of Yukalov and Kleinert's paper~\cite{YK}. The most successful effective theory is due to Bogoliubov~\cite{Bog}, \cite{Bog1}, \cite{Bog2}. Based on the hypothesis that BEC takes place in the ground state, Bogoliubov introduced some approximations in order to turn the problem into soluble. He dropped part of the Hamiltonian (Bogoliubov truncation, BT), and replaced $a_0^*$ and $a_0$ by volume ($V$) dependent complex numbers (Bogoliubov substitution, BS). The upshot of the approximations is a Hamiltonian that can be diagonalized by Bogoliubov's canonical transformation. In the subsequent physics literature BT and BS appear either separately or coupled together, with an embarrassing variety of truncation schemes. In the past and also more recently much effort has been made to justify these approximations. Their status from the point of view of mathematical rigor and also of physical significance proved to be quite different.

In a fundamental paper \cite{Gin} Ginibre studied BS alone and proved that
if the pressure is maximized with respect to the complex number substituting $a_0$, the approximate
pressure converges to the true one in the thermodynamic limit. This still did not imply
that BS was indeed "exact", because concerning the
states, it represents a drastic reduction of Fock space: only states of the form
$\widehat{z}\psi$ are considered, where $z$ is a complex number, $\widehat{z}=e^{za_0^*-\overline{z}a_0}$ and $\psi$ is an element of the Fock-subspace $F'$ built
on the plane waves $\{e^{i\k\cdot\r}/\sqrt{V}\}_{\k\neq 0}$.
Justification of BS means therefore to prove that along with the pressure the averages of the physically most relevant operators are also correctly reproduced by it. A step in this direction was made in two
recent papers \cite{LSY}, \cite{Su}. These showed that if $a_0\to z(V)$ is a
complex number substitution that makes the pressure of the BS model converge to the true
pressure in the thermodynamic limit, then
$|z(V)|^2/V$ and the exact $\langle a_0^*a_0\rangle/V$ have the same limits as $V$ tends
to infinity. The result implies that BS predicts BEC, when it occurs,
with the right density of condensate.

No similar rigorous justification exists for any kind of truncation of the Hamiltonian.
While BS alone is quantitatively correct in the sense described above, BT cannot be hoped to be quantitatively correct except for the cases of extremely low densities and weak interactions. Even qualitative correctness can partly be jeopardized, as it happens for instance in relation with the gap. Truncation of the Hamiltonian changes the ground state energy in the order of the volume and introduces unphysical features as the non-locality of the interaction. By this, it causes the breakdown of the local gauge invariance of the field operators in coordinate space, and changes the global $U(1)$ symmetry into local one in $\k$-space. Therefore, truncated Hamiltonians should be considered as separate models that can describe certain but not all expected properties of the physical system.

There exists a particular class of truncated Hamiltonians in which truncation is variational in the sense that the eigenvalues are upper bounds to the lowest eigenvalue of the physical Hamiltonian in every eigensubspace of the total momentum. Such a family of Hamiltonians appeared fifty years ago in a paper by Valatin and Butler~\cite{VB}. The Valatin-Butler (VB) Hamiltonians depend on the parameters of Bogoliubov's canonical transformation and are diagonal in the quasiparticle creation and annihilation operators obtained through this transformation. The corresponding quasiparticle vacuum state contains pairs of physical particles of opposite momenta. The excitation of such pairs was thought already by Bogoliubov to be the dominant effect of the interaction on the ground state of a Bose system. Valatin and Butler therefore expected to obtain a good quasiparticle theory by minimization of the energy of the vacuum state over the set of parameters of the Bogoliubov transformation and by retaining the diagonal Hamiltonian defined by the minimizing parameters. They failed in this program, but their paper is remarkable and to locate the source of their failure even fifty years later seems to us relevant. Our conclusion, based on the analysis of Section III and valid for positive pair interactions with a positive Fourier transform, can be summarized in two points. First, within the family of eigenstates of the VB Hamiltonians the state of lowest energy is not a vacuum state but a state with $\sim\sqrt{V}$ quasiparticles of zero momentum. Second, Bogoliubov's transformation in the zero mode leads to anomalous fluctuations of the number of condensed physical particles, increasing the lowest attainable energy.

The two problems of the VB approach are related and both disappear if Bogoliubov's transformation in the zero mode is dropped and replaced by Bogoliubov's substitution. This was first done by Nozi\`eres and Saint James~\cite{NStJ}. In Section IV we analyze in detail and somewhat extend the Nozi\`eres-Saint James (NStJ) approach. First, BS is replaced by a shift, $a_0-z$, according to Araki and Woods~\cite{AW}. This permits us to keep the zero mode in consideration and to allow deviations from a coherent state. We then introduce variationally truncated Hamiltonians defined now on the full Fock space and depending on the parameters of Bogoliubov's transformation for $\k\neq 0$ and on $z$ for $\k=0$, consider the family of their eigenstates, and prove that the lowest energy is attained on a vacuum state with an energy density lower than the VB minimum. Finally, we add a Bogoliubov transformation also in the zero mode and minimize the energy of the vacuum state with respect to it. While the gain in energy is only of the order of unity, by this way the gap of the associated quasiparticle theory, discussed in a moment, can be made continuous at $\k=0$.

Although we are interested only in minimizing the energy of our trial functions, because of their appearance as eigenstates of quasiparticle Hamiltonians we compute the corresponding gaps and comment on them at several places. In the spectrum of a physical Hamiltonian with a translation invariant and not too slowly decaying interaction there is no gap above the ground state (in fact, there is a gap that disappears as $V$ tends to infinity). This was first shown by Hugenholtz and Pines~\cite{HP} for $\k=0$, then by Bogoliubov~\cite{Bog2} for $\k\to 0$ and somewhat later a rigorous proof, related with the breakdown of Galilei invariance and not with BEC, was given by Swieca~\cite{Swi}. On the other hand, truncated models most often produce a gap to excitations. Sometimes the gap disappears for $\k=0$ but stays for $\k\to 0$ (see \cite{Su2} and Section III), sometimes it stays for both and the two gaps can be different (Sections III.B and IV.B) or can coincide (Section IV.C). It is because of this gap that the work of Girardeau and Arnowitt \cite{GA} was considered later to be inconsistent \cite{HP}, although they minimized correctly the energy of their trial state. (Working with a fixed number of particles, they avoided the problem of anomalous fluctuations.) However, a gap does not mean the \emph{internal} inconsistency of an approximate theory. The quasiparticle theories emerging as a by-product from the present variational treatment are, in fact, consistent by themselves. They can be gapful without contradicting Goldstone's theorem applied to \emph{them} because there is no BEC of \emph{quasiparticles} in their ground state.

\newsec{Preliminaries}

We are going to perform a thorough study of variational wave functions of the product form, factorizing according to the wave vectors, for the ground state of the Hamiltonian
\bea\label{H}
H=\sum\veps(\k) N_k -\mu N
+\frac{1}{2V}\sum_{\q,\k,\k'}v(\q)
a^*_{\k+\q}a^*_{\k'-\q}a_{\k'}a_{\k}\nonumber\\
\eea
Here $a^*_{\k}$ and $a_\k$ create, respectively, annihilate a boson in the single-particle state $(1/\sqrt{V})\exp\{i\k\cdot\r\}$, where $\k=(2\pi/L)(n_1,n_2,n_3)$ with $n_i$ integers and $L^3=V$, $\veps(\k)=\hbar^2\k^2/2m$, $\mu$ and $v(\q)=v(-\q)$ are real numbers, $N_\k=a^*_{\k}a_{\k}$ and $N=\sum N_\k$. $H$ is defined with periodic boundary conditions on a cube of side $L$; $v$ is the Fourier transform of an integrable pair potential $\varphi$,
\be
v(\k)=\int\varphi(\r)e^{-i\k\cdot\r}\d\r,
\ee
and most of the time both $v$ and $\varphi$ will be supposed to be nonnegative. A typical example of a positive interaction with a positive Fourier transform is a Gaussian interaction. More examples can be obtained by convoluting a real nonnegative function $\psi(\x)$ with $\psi(-\x)$: if
\be
\varphi(\x)=\int\psi(\y)\psi(\y+\x)\d\y,
\ee
then both $\varphi$ and $v=\hat\varphi=|\hat\psi|^2$ are nonnegative. Note that the chemical potential is included in the definition of $H$, so the ground state of $H$ is understood in Fock space. For $\varphi\geq 0$ the interesting region of the chemical potential is $\mu>0$; for $\mu\leq 0$ the ground state is the vacuum.

\subsection{Pairing product states}

Pairing product states are wave functions of the general form $\Phi=\widehat{\Phi}|0\rangle$ where $|0\rangle$ is the physical vacuum state and
\be\label{BCS}
\widehat{\Phi}=\left(\sum_{m=0}^\infty\beta_{0,m}(a_0^*)^{m}\right)\prod_{\{\k,-\k\}\neq\{0\}}
\left(\sum_{m=0}^\infty\beta_{\k,m}(a_\k^*a_{-\k}^*)^{m}\right)
\ee
A variational ansatz of this kind suggests a division of $H$ in three parts,
\be
H=H_0+H_{\neg 0}+H_R
\ee
where $H_0$ collects all terms containing $N_0$, $N_0^2$, $A_0=a_0^2$ and $A_0^*$, $\Hnot0$ collects all terms containing $N_\k$, $N_\k^2$, $A_\k=a_\k a_{-\k}$ and $A_\k^*$ \emph{only} for $\k\neq 0$, and $H_R$ is the sum of the remaining terms of $H$ not contributing to the energy of $\Phi$: $(\Phi,H_R\Phi)=0$. Thus,
\bea\label{H0}
H_0&=&-\mu N_0+\frac{v(0)}{2V}(N_0^2-N_0)+\frac{N_0}{V}\sum_{\k\neq 0}[v(0)+v(\k)]N_\k\nonumber\\
&+&\frac{1}{2V}\sum_{\k\neq 0}v(\k)[A_0^*A_\k+A_\k^*A_0]
\eea
and
\bea\label{Hnot0}
\Hnot0=\sum_{\k\neq 0}[\veps(\k)-\mu]N_\k+\frac{v(0)}{2V}\sum_{\k\neq 0}(N_\k^2-N_\k)\nonumber\\
+\frac{1}{2V}\sum_{0\neq\k\neq\k'\neq 0}[v(0)+v(\k-\k')]N_\k N_{\k'}\nonumber\\
+\frac{1}{2V}\sum_{\k\neq 0}\sum_{\k'\neq 0,\pm\k}v(\k-\k')A_{\k'}^*A_\k\ .
\eea
For an operator $B$ let $\langle B\rangle=(\Phi,B\Phi)$. In $\langle H_0\rangle$ and $\langle \Hnot0\rangle$ the mean value of $A_{\k'}^*A_\k$ factorizes. This makes it possible to choose $\langle A_\k\rangle$ independent complex variables: prescribing any set $\{z_\k\}$ of complex numbers, $\Phi$ can be chosen so that $\langle A_\k\rangle=z_\k$ for all $\k$. The Valatin-Butler vacuum state with complex $g_\k$ serves below as an example. Define
\bea\label{nu12}
\nu_1&=&\frac{1}{V}\sum_{\k\neq 0}[v(0)+v(\k)]\langle N_\k\rangle\nonumber\\
\nu_2&=&\frac{1}{V}\sum_{\k\neq 0}v(\k){\rm Re\,}\langle A_\k\rangle\ .
\eea
Choose $\langle A_0\rangle$ to be real and nonnegative. When minimizing $\langle H\rangle$, the sign of $\nu_2$ has no influence on the value of $\langle\Hnot0\rangle$ but $\nu_2\leq 0$ is necessary to minimize $\langle H_0\rangle$. We then find
\bea\label{avH0}
\langle H_0\rangle=&-&\left(\mu-\nu_1+|\nu_2|+\frac{v(0)}{2V}\right)\,\langle N_0\rangle+\frac{v(0)}{2V}\langle N_0\rangle^2\nonumber\\
&+&\frac{v(0)}{2V}D^2(N_0)+|\nu_2|\,[\langle N_0\rangle-\langle A_0\rangle]
\eea
with
\be
D^2(N_0)=\langle N_0^2\rangle-\langle N_0\rangle^2\,.
\ee
Observe now that
\be
\langle N_0\rangle=\|a_0\Phi\|^2\ ,\quad \langle N_0\rangle+1=\|a_0^*\Phi\|^2
\ee
and therefore by Schwarz inequality
\be
\langle A_0\rangle\leq \langle N_0\rangle\sqrt{1+1/\langle N_0\rangle}< \langle N_0\rangle + \frac{1}{2}\ .
\ee
Because $D^2(N_0)\geq 0$, we obtain the following lower bound on $\langle H_0\rangle$:
\be\label{H0>}
\langle H_0\rangle\geq -\left(\mu-\nu_1+|\nu_2|+\frac{v(0)}{2V}\right)\,\langle N_0\rangle
+\frac{v(0)}{2V}\langle N_0\rangle^2-\frac{|\nu_2|}{2}\ .
\ee
The minimum in $\langle N_0\rangle$ of this bound is either $-|\nu_2|/2$ if $\mu-\nu_1+|\nu_2|<0$ or
\be\label{lb}
\langle H_0\rangle\geq -\frac{V}{2v(0)}(\mu-\nu_1+|\nu_2|)^2-\frac{1}{2}(\mu-\nu_1+2|\nu_2|)+O(V^{-1})
\ee
obtained with
\be\label{expN0}
\langle N_0\rangle=\frac{V}{v(0)}(\mu-\nu_1+|\nu_2|)+\frac{1}{2}
\ee
if $\mu-\nu_1+|\nu_2|\geq 0$. A particular case of (\ref{BCS}) is when $\Phi$ is an eigenstate of $N_0$, so among all $\beta_{\k,m}$ only $\beta_{0,m}$ is non-vanishing for a single value of $m$. Then $\langle N_\k\rangle=\langle A_\k\rangle=0$ for $\k\neq 0$, $\langle A_0\rangle=D^2(N_0)=0$ and
\be
\langle H\rangle=\langle H_0\rangle=-\left(\mu+\frac{v(0)}{2V}\right)m+\frac{v(0)}{2V}m^2
\ee
leaving $m$ as a single variational parameter. Minimization then yields
\be\label{EMF}
\langle H\rangle=E_{\rm MF}=-\frac{\mu^2V}{2v(0)}-\frac{\mu}{2}+O(V^{-1})
\ee
obtained with $m=\mu V/v(0)+1/2$. This is the ground state energy of the mean field and also of the so-called perturbed mean field or full diagonal models \cite{DLP,DMP,Su2}. It will serve as a reference value, the parameters will have to be chosen so as to obtain an energy lower than $E_{\rm MF}$, by getting the closest possible to the lower bound (\ref{lb}) with the largest possible value of $|\nu_2|-\nu_1$ without loosing too much energy in $\langle\Hnot0\rangle$.

\subsection{Coherent product states}

The ansatz (\ref{BCS}) will be partly restricted, partly extended. We introduce two sets of quasiparticle annihilation and creation operators,
\be\label{shift}
\tilda_\k=a_\k-z_\k ,\qquad\tilda_\k^*=a_\k^*-\overline{z_\k}
\ee
and
\bea\label{bogotrans}
b_\k&=&\frac{1}{\sqrt{1-|g_\k|^2}}(\tilda_\k-g_\k \tilda_{-\k}^*)\nonumber\\
b_\k^*&=&\frac{1}{\sqrt{1-|g_\k|^2}}(\tilda_\k^*-\overline{g_\k}\, \tilda_{-\k})
\eea
where $z_\k$ and $g_\k$ are complex numbers, $|g_\k|<1$ and $g_\k=g_{-\k}$. The coherent state
\be
Z=\exp\left[\sum_{\bf k}(z_{\bf k}a_{\bf k}^*-\overline{z_{\bf k}}a_{\bf k})\right]|0\rangle
\ee
is the common vacuum of each $\tilda_\k$, $\tilda_\k Z=0$. It is then straightforward to verify that
\be\label{dressed}
\Phi_0=\prod_\k(1-|g_\k|^2)^{\frac{1}{4}}\exp\left[{\frac{1}{2}g_\k\tilda_\k^*\tilda_{-\k}^*}\right]Z
\ee
is the vacuum state of each $b_\k$, i.e. $b_\k\Phi_0=0$. On $\Phi_0$ one can build up an orthonormal basis in the Fock space,
\be\label{basis}
\Phi_{\{m_\k\}}=\prod_\k\frac{1}{\sqrt{m_\k!}}(b_\k^*)^{m_\k}\Phi_0
\ee
by letting $\{m_\k\}$ run over terminating sequences of nonnegative integers. Unless $z_\k=0$ for all $\k\neq 0$, $Z$ and thus $\Phi_{\{m_\k\}}$ are not eigenstates of the total momentum. Choosing $|z_{-\k}|=|z_\k|$ one has at least
\be
\left(\Phi_{\{m_\k\}},\left(\sum\k N_\k\right)\Phi_{\{m_\k\}}\right)=\sum_\k  m_\k\k\ .
\ee

Given a basis (\ref{basis}), it is possible to define a quasiparticle theory of the interacting Bose gas via the Hamiltonian
\be\label{QPHam}
H_{\rm QP}=\sum_{\{m_\k\geq 0\}}E_{\{m_\k\}}|\Phi_{\{m_\k\}}\rangle\langle\Phi_{\{m_\k\}}|
\ee
where
\be\label{energy}
E_{\{m_\k\}}=\left(\Phi_{\{m_\k\}},H\Phi_{\{m_\k\}}\right).
\ee
Such a theory has the merit of being variational to the ground state of $H$. If $z_\k=0$ for $\k\neq 0$ and, hence, $H_{\rm QP}$ commutes with the total momentum operator, the quasiparticle theory is variational also in each momentum eigenspace,
\be\label{Eineq}
E_{\{m_\k\}}\geq \inf{\rm spec}\ H\Pi_{\{m_\k\}}
\ee
where $\Pi_{\{m_\k\}}$ projects onto the subspace of states of momentum $\hbar\sum m_\k \k$. Those concerned with effective theories of the interacting Bose gas may not optimize $H_{\rm QP}$ by minimizing its ground state energy, but try instead to make the gap above the ground state disappear. In this paper we insist on an unbiased minimization of the energy, yielding what we call the \emph{quasiparticle ground state} (QPGS) of $H$. At several places we shall comment on the gap in the spectrum of $H_{\rm QP}$.

If $z_\k=0$ for all $\k\neq 0$, the dressed vacuum state (\ref{dressed}) and states (\ref{basis}) with  $m_{-\k}=m_\k$ are also pairing. If $z_\k\neq 0$ for some $\k\neq 0$, $\Phi_{\{m_\k\}}$ provides a nonvanishing average of $H_R$ and, thus, is not of the type (\ref{BCS}). Such product states can have lower energies but not necessarily in the order of the volume. One special case of them, when all $g_\k=0$, can immediately be excluded. Indeed, a simple calculation shows that whenever $v\geq 0$, $\min(Z,HZ)=-\mu^2V/2v(0)>E_{\rm MF}$, obtained with $z_\k=\sqrt{\mu V/v(0)}\delta_{\k,0}$. With the benefit of hindsight one can state that for $\k=0$ the shift (\ref{shift}) is more important than the Bogoliubov transformation (\ref{bogotrans}) while for $\k\neq 0$ the opposite holds true. A full variational treatment, extending to the sets $\{g_\k\}$, $\{z_\k\}$ and $\{m_\k\}$ seems at any rate too complicated. We will therefore focus on the case $z_\k=z\delta_{\k,0}$. The variational parameters are then $z$ and $g_\k$ (including $g_0$), and all basis functions (\ref{basis}) will be considered. Two important special cases are $z=0$, the Valatin-Butler (VB) scheme and $g_0=0$, the Nozi\`eres-Saint James-Araki-Woods (NStJAW) scheme.

\newsec{The Valatin-Butler scheme}

Valatin and Butler~\cite{VB}, inspired by the then fresh BCS paper~\cite{BCS}, proposed a quasiparticle theory for bosons in the lines described above, with $\tilda_\k=a_\k$ (all $z_\k=0$) and $g_\k$ real. The dressed vacuum (\ref{dressed}) is then pairing also for $\k=0$. In choosing $g_\k$ real, Valatin and Butler followed Bogoliubov~\cite{Bog1}. In principle, this choice restricts generality since the most general gauge transformation $a_\k\to e^{i(\t\cdot\k+\alpha)}a_\k$ that leaves $H$ invariant for all $\t\in\R^3$ and $\alpha\in\R$ is not enough to eliminate an arbitrary $\k$-dependent complex unit factor of $g_\k$. However, the ground state of $H$ is unique and one may suppose that its QPGS is also unique. Then $g_\k$ can indeed be chosen to be real.

Valatin and Butler made some \emph{ad hoc} assumptions which eventually falsified the minimization of the energy. Their condition on the interaction was vague; they thought it to be repulsive but, most important, such that the QPGS is the vacuum for some $\{g_\k\}$. This assumption was inconsistent with the variational determination of $g_0$, so they chose it to make the gap to $\k=0$ excitations vanish. However, because
\be
(\Phi_0,N_0\Phi_0)=\frac{g_0^2}{1-g_0^2}\ ,
\ee
it was in fact the condensate density that they adjusted to make the gap disappear. This was a strongly contestable step for at least two reasons. First, the gap in question is in the spectrum of $H_{\rm QP}$, therefore the breakdown of a continuous symmetry of $H_{\rm QP}$ could only be in conflict with it. If the ground state of $H_{\rm QP}$ is the vacuum state, a gap above it does not contradict Goldstone's theorem applied to $H_{\rm QP}$. One should be worried only if $H_{\rm QP}$ had a ground state with a macroscopic expectation value of $b_0^*b_0$ and still there would be a gap above the ground state. Second, the density of the condensate of physical particles can be varied to minimize the energy, but not for other purposes. The only free physical parameter is the chemical potential $\mu$. If the $N$-particle interaction energy is superstable (meaning that in any finite volume $V$ it eventually increases with $N$ not slower than $N^2/V$), the full particle density is a strictly increasing function of $\mu$ (at zero temperature for $\mu\geq 0$, if the interaction is nonnegative). Therefore, instead of $\mu$ one may freely prescribe the full density, but not that of the condensate.

The explicit form of the quasiparticle Hamiltonian (\ref{QPHam}) can be found e.g. by rewriting $H$ in terms of the operators $b_\k$ and $b_\k^*$ and keeping only those terms commuting with each $M_\k=b_\k^*b_\k$. Let
\be\label{h-chi}
h_\k=\frac{g_\k^2}{1-g_\k^2}=(\Phi_0,N_\k\Phi_0),\quad
\chi_\k=\frac{g_\k}{1-g_\k^2}=(\Phi_0,A_\k\Phi_0)
\ee
so that $\chi_\k^2=h_\k(h_\k+1)$. A lengthy but straightforward calculation yields
\begin{widetext}
\be\label{Hg}
H_{\rm QP}=w_0+\sum_\k e_\k M_\k+\sum_\k w_{\k\k}(M_\k^2-M_\k)+\sum_{\k\neq\k'}w_{\k\k'}M_\k M_{\k'}
\ee
where
\be\label{w0}
w_0=\sum_\k(\veps(\k)-\mu)h_\k+\frac{v(0)}{2V}\left(\sum_{\k}h_{\k}\right)^2 +\frac{1}{2V}\sum_{\k,\k'}v(\k-\k')
(\chi_\k\chi_{\k'}+h_\k h_{\k'})\ ,
\ee
\be\label{ek}
e_\k=(1+2h_\k)\left(\veps(\k)-\mu+\frac{1}{V}\sum_{\k'}\,[v(0)+v(\k-\k')]\,h_{\k'}\right)
+2\chi_\k\frac{1}{V}\sum_{\k'}v(\k-\k')\chi_{\k'} ,
\ee
\be
w_{\k\k}=\frac{v(0)}{2V}(1+2h_\k)^2+\frac{v(2\k)}{V}\chi_\k^2\ ,
\ee
\be\label{wkk'}
w_{\k\k'}=\frac{1}{2V}[v(0)+v(\k-\k')]+\frac{1}{2V}[2v(0)+v(\k+\k')+v(\k-\k')](h_\k+h_{\k'}+2h_\k h_{\k'})
+\frac{1}{V}[v(\k+\k')+v(\k-\k')]\chi_\k\chi_{\k'}\ .
\ee
\end{widetext}
It is seen that $e_{-\k}=e_\k$ and $w_{-\k,-\k'}=w_{\k\k'}=w_{\k'\k}$ for all $\k,\k'$. From Eq.~(\ref{Hg}) the energies (\ref{energy}) can be obtained by replacing the operators $M_\k$ by the integers $m_\k$. Alternatively, one can take the average with $\Phi_{\{m_\k\}}$ of Eqs.~(\ref{H0}) and (\ref{Hnot0}) and use
\be\label{avfirst}
\langle N_\k\rangle=h_\k+(1+h_\k)m_\k+h_\k m_{-\k},
\ee
\be
\langle A_\k\rangle=\chi_\k(1+m_\k+m_{-\k}),
\ee
\bea\label{avlast}
D^2(N_\k)&\equiv&\langle N_\k^2\rangle-\langle N_\k\rangle^2\nonumber\\
&=&\langle N_\k N_{-\k}\rangle-\langle N_\k\rangle\langle N_{-\k}\rangle\nonumber\\
&=&\chi_\k^2(1+\delta_{\k,0}+m_\k+m_{-\k}+2m_\k m_{-\k}).\nonumber\\
\phantom{a}
\eea
The expressions for $w_0$ and $e_\k$ were already given in \cite{VB}. A particular member of the family of operators (\ref{Hg}), obtained with all $g_\k=0$, is the perturbed mean field or full diagonal (FD) Hamiltonian which has long been serving to test different ideas concerning the interacting Bose gas \cite{Su2}, \cite{DLP}, \cite{DMP}. It will prove to be useful here as well. In a more convenient form it reads
\bea\label{HFD}
H_{\rm FD}=\sum\veps(\k) N_k-\mu N+\frac{v(0)}{2V}(N^2-N)\nonumber\\
+\frac{1}{2V}\sum_{\k\neq\k'}v(\k-\k')N_\k N_{\k'}\ .
\eea
It is clear that the overall minimum of $E_{\{m_\k\}}(\{g_\k\})$ cannot be higher than the ground state energy of $H_{\rm FD}$. Suppose that $v(\k)\geq 0$ and is continuous at $0$ with $v(0)>0$. Then for $\mu\leq 0$ the ground state of $H_{\rm FD}$ is the physical vacuum with zero energy. If $\mu>0$, it is $(m_0!)^{-\frac{1}{2}}(a_0^*)^{m_0}|0\rangle$ with 
\be
m_0=\mu V/v(0)+\frac{1}{2}\equiv m_V. 
\ee
(We may suppose that $m_V$ is an integer.) This gives
\be\label{E00}
E_{m_0=m_V}=-\frac{\mu^2V}{2v(0)}\left(1+\frac{v(0)}{2\mu V}\right)^2=E_{\rm MF}
\ee
for the ground state energy of $H_{\rm FD}$, cf. Eq.~(\ref{EMF}). What about the gap to excitations? If $\mu\leq 0$, the gap is at least $|\mu|$ to any excitation of the vacuum. If $\mu>0$, because
$m_0=m_V$ was chosen to minimize the quadratic polynomial $-\mu m_0+[v(0)/2V](m_0^2-m_0)$, we find
\be\label{gap00}
E_{m_0=m_V\pm n}-E_{m_0=m_V}=\frac{v(0)}{2V}n^2
\ee
which for $n=o(\sqrt{V})$ tends to zero as $V$ goes to infinity. Thus, the gap to $\k=0$ excitations vanishes, and this is consistent with the breakdown of the gauge invariance during BEC in this model system \cite{Su2}. On the other hand, for $\k\neq 0$
\bea\label{gap0k}
\lefteqn{
E_{m_0=m_V,m_\k=1}-E_{m_0=m_V}}\nonumber\\
&&=\veps(\k)-\mu+\frac{v(0)+v(\k)}{V}m_V\geq\mu\frac{v(\k)}{v(0)}\ .
\eea
(Above it is understood that $m_{\k'}=0$ unless otherwise stated.) So there remains a partial gap to $\k\neq 0$ excitations in the limit of infinite volume, which tends to $\mu$ with $\k$ going to zero.

\subsection{The quasiparticle ground state in the VB scheme}

\emph{Proposition 1.} Let the pair interaction be nonnegative, $\varphi\geq 0$. Then the quasiparticle ground state is not a dressed vacuum state. If also $v>0$, the energy difference between any vacuum state and the QPGS is of the order of the volume.

\emph{Proof.} Note that $(\Phi_0,H\Phi_0)=w_0$. $\varphi\geq 0$ implies that $v$ is positive definite, that is,
\[
\sum_{\k,\k'}v(\k-\k')\overline{\xi_\k}\,\xi_{\k'}\geq 0
\]
for any choice of $\xi_\k$. Therefore, even without supposing $v(\k)\geq 0$,
\bea\label{w0bis}
w_0=\sum_\k\veps_\k h_\k
+\frac{1}{2V}\sum_{\k,\k'}v(\k-\k')(\chi_\k\chi_{\k'}+h_\k h_{\k'})\nonumber\\
+\frac{v(0)}{2V}\left(\sum_{\k}h_{\k}-\frac{\mu V}{v(0)}\right)^2
-\frac{\mu^2V}{2v(0)}>-\frac{\mu^2V}{2v(0)}\ .
\eea
If $v\geq 0$, the lower bound is slightly above the ground state energy (\ref{E00}) of the FD model. Otherwise, using $(m_V!)^{-\frac{1}{2}}(a_0^*)^{m_V}|0\rangle$ as a variational wave function, one can see that the ground state energy of $H_{\rm FD}$ is bounded above by the value (\ref{E00}), so is certainly smaller than $w_0$. This proves that the QPGS is not a vacuum state. Assume now $\varphi\geq 0$ and $v>0$. A lower bound on $w_0$ is obtained by dropping $$\sum_{\k,\k'}v(\k-\k')\chi_\k\chi_{\k'}\geq 0$$
from the right-hand side of Eq.~(\ref{w0}). Let $k_\mu=\hbar^{-1}\sqrt{2m\mu}$. If $k=|\k|\geq k_\mu$ then $\veps(\k)-\mu\geq 0$. The other terms for $k$ or $k'\geq k_\mu$ being also positive, the lower bound on $w_0$ can further be decreased by restricting the summations to $k,k'<k_\mu$. Dropping also $\sum\veps(\k) h_\k\geq 0$, we find
\bea\label{lowerw0}
w_0&\geq& -\mu x+\frac{v(0)}{2V}x^2
+\frac{1}{2V}\sum_{k,k'<k_\mu}v(\k-\k')h_\k h_{\k'}\nonumber\\
&\geq& -\mu x+\frac{1}{2V}\left[v(0)+\min_{k<2k_\mu}v(\k)\right]x^2
\eea
where $x=\sum_{k<k_\mu}h_\k$. Minimizing over $x$,
\be
\frac{w_0}{V}\geq -\frac{\mu^2}{2[v(0)+\min_{k<2k_\mu}v(\k)]}>-\frac{\mu^2}{2v(0)}\ ,
\ee
obtained by inserting
\[
x=\frac{\mu V}{v(0)+\min_{k<2k_\mu}v(\k)}
\]
in (\ref{lowerw0}). Comparison with (\ref{E00}) shows that the energy \emph{density} of any vacuum state is higher than the energy density of the ground state of the full diagonal model and, therefore, of the energy density of the QPGS.

\emph{Proposition 2.} Let $v\geq 0$. Suppose that the QPGS is unique. Then in the QPGS only $m_0$ can differ from zero.

\emph{Proof.} Uniqueness implies $m_{-\k}=m_\k$ for all $\k$. Suppose $m_{-\k}=m_\k=n>0$ for some $\k\neq 0$. Compare the energy $E_{m_{-\k}=m_\k=n}$ of this state with the energy $E_{m_{-\k}=0,m_\k=2n}$ of the state obtained by particle transfer from $-\k$ to $\k$ while keeping $m_{-\k'}=m_{\k'}$ unchanged for $\k'\neq\pm\k$. To compute the difference of these energies use
\be
\sum_{\k'\neq\pm\k}w_{\k',-\k}m_{\k'}=\sum_{\k'\neq\pm\k}w_{-\k',-\k}m_{-\k'}
=\sum_{\k'\neq\pm\k}w_{\k',\k}m_{\k'}\ ,
\ee
and
\be\label{deltaw}
w_{\k,-\k}-w_{\k\k}=\frac{v(2\k)}{2V}+\frac{1}{V}[2v(0)+v(2\k)]\chi_\k^2>0.
\ee
With them,
\bea\label{deltaE}
E_{m_{-\k}=m_\k=n}-E_{m_{-\k}=0,m_\k=2n}\nonumber\\
=2(w_{\k,-\k}-w_{\k\k})n^2\geq\frac{v(2\k)}{V}n^2\geq 0
\eea
Thus the starting state is either not the QPGS or it is one but not the only one of them, contradicting uniqueness.

Note that not only $w_{\k,\pm\k}$ are positive but in fact all $w_{\k,\k'}>0$. From Eq.~(\ref{wkk'}) one easily deduces
\be\label{wineq}
w_{\k,\k'}\geq\frac{v(0)+v(\k-\k')}{2V}+\frac{v(0)}{V}(h_\k+h_{\k'}+2h_\k h_{\k'})>0\ .
\ee
For (\ref{deltaw}) and (\ref{wineq}) to hold, only $v\geq 0$ is necessary.

\emph{Theorem 1.} Let $\varphi\geq 0$ and $v\geq 0$. Consider the basis (\ref{basis}) with $z_\k\equiv 0$. For $\mu>0$ the QPGS is either the ground state of the full diagonal model or a small perturbation of it having the same energy density, or an eigenstate of $H_{\rm QP}$ with
\bea\label{root}
\chi_0&\propto&\sqrt{V}\ \mbox{ and }m_0\propto\sqrt{V}\ \mbox{as}\ V\to\infty,\nonumber\\
\chi_\k&=&O(1)\ \mbox{for}\ \k\neq 0\ ,\mbox{ and } \chi_0\nu_2<0
\eea
and with an energy density smaller than $-\mu^2V/2v(0)$. If the decay of $v(\k)$ is slow enough ($\varphi$ is sharp enough) then for small $\mu$ the last option holds.

\emph{Example.} If
\be\label{example}
\varphi(\r)=(4\pi\lambda^2)^{-\frac{3}{2}}v(0)\,e^{-r^2/4\lambda^2},
\quad v(\k)=v(0)e^{-(\lambda k)^2}
\ee
then the condition of sharpness is $\lambda<a_B/363$ where $a_B=v(0)m/4\pi\hbar^2$, the scattering length in Born approximation.

\emph{Remarks.} (i) Because the average of $N_0$ in a pairing product state ($z_0=0$) is
\be
\langle N_0\rangle=h_0+m_0+2h_0 m_0,
\ee
$\langle N_0\rangle\propto V$ in all cases. Thus, in the QPGS there is BEC of \emph{physical} particles.

\noindent
(ii) When (\ref{root}) holds for the QPGS, there is no macroscopic and no generalized condensation for \emph{quasiparticles}:
$\langle M_0\rangle=m_0\propto\sqrt{V}$ and
$\langle M_\k\rangle=m_\k=0$ for $\k\neq 0$. Moreover, the interaction being nonlocal, Swieca's proof \cite{Swi} does not apply either. Therefore there is no reason to expect that the gap above the ground state of $H_{\rm QP}$ disappears. The gap will be derived in the next section.

\noindent
(iii) Of the two conditions $v\geq 0$ is the more important. $\varphi\geq 0$ serves only to guarantee that the QPGS is not a vacuum state. It seems to be difficult to find a superstable interaction whose QPGS in the VB scheme \emph{is} a vacuum state.

\emph{Proof of the theorem.} From the two propositions above, the quasiparticle ground state energy is of the form
\be\label{Egs}
E_{m_0}=w_0+e_0 m_0+w_{00}(m_0^2-m_0).
\ee
Because the last term is nonnegative, this can be smaller than the vacuum energy $w_0$ only if $e_0$ is negative. The minimum of (\ref{Egs}) is attained with
\be
m_0=-\frac{e_0}{2w_{00}}+\frac{1}{2}\equiv m_V
\ee
that we suppose to be an integer, and its value is
\be\label{Eg0}
E_{m_0=m_V}=w_0-\frac{1}{4w_{00}}(w_{00}-e_0)^2=w_0-w_{00}m_V^2
\ee

We minimize $E_{m_0=m_V}$ with respect to $\chi_0$ which, in contrast to $g_0$, can take on any real value. For further use we introduce
\be\label{xk}
x_\k=\frac{\d h_\k}{\d\chi_\k}=\frac{2\chi_\k}{1+2h_\k}\in(-1,1).
\ee
Let
\be
C=\frac{V}{v(0)}\,\nu_2\ ,\qquad D=\frac{V}{v(0)}(\mu-\nu_1).
\ee
Here and throughout the rest of the paper $\nu_1$ and $\nu_2$ are given by Eq.~(\ref{nu12}) with $\langle N_k\rangle=h_\k$ and $\langle A_\k\rangle=\chi_\k$:
\be\label{nu12-final}
\nu_1=\frac{1}{V}\sum_{\k\neq 0}[v(0)+v(\k)]h_\k,\qquad
\nu_2=\frac{1}{V}\sum_{\k\neq 0}v(\k)\chi_\k\ .
\ee
After some algebra one finds
\bea
G_V(x_0)=\frac{V}{v(0)}\sqrt{1-x_0^2}\left(1+\frac{1}{2}x_0^2\right)^2\frac{\partial E_{m_0=m_V}}{\partial\chi_0}\nonumber\\
=(D+1-Cx_0)(1-x_0^2)^2[(D+1)x_0+2C]\nonumber\\
+\frac{3}{4}x_0\left(1+\frac{1}{2}x_0^2\right)^2.
\eea
Since the second term is of order 1 while the first is typically of order $V^2$, the zeros of $G_V$ are in a distance $O(V^{-2})$ to those of the first term. The first factor, $D+1-Cx_0$ must be positive, otherwise $e_0>0$. Thus, $G_V$ has three zeros in distance $O(V^{-2})$ to $\pm 1$ and $-2C/(D+1)$, respectively, and only in one of these points is the energy minimal. To minimize $E_{m_0=m_V}$ one has to maximize $m_V$, implying $\nu_2 x_0\leq 0$, see Eq.~(\ref{mV}) below. For example, if $\nu_2\leq 0$ then $x_0\geq 0$, so the root we look for is either close to $|2C/(D+1)|$ or to 1. Now $|2C/(D+1)|$ may or may not be smaller than 1 and when smaller, it is the right choice. To see this note that if $\chi_\k=0$ for $\k\neq 0$ then $C=0$ and $G_V(0)=0$. Thus, at $\chi_0=0$ the energy is stationary and in fact minimum with value $E_{\rm MF}$. Indeed, $\langle\Hnot0\rangle=0$ in this case, therefore $\langle H\rangle=\langle H_0\rangle\geq E_{\rm MF}$, see (\ref{Hnot0}), (\ref{lb}), (\ref{EMF}), and the lower bound is attained with the full diagonal model of ground state energy $E_{\rm MF}$. If $\chi_\k$ are nonvanishing and $|2C/(D+1)|\leq 1$ then by continuity it is still $x_0=-2C/(D+1)+o(1)$ that defines the minimum, and this remains true (with a negative $o(1)$ correction) also if $|2C/(D+1)|$ converges to 1 from above as $V$ tends to infinity.

\emph{Lemma 1.} Under the condition that $x_0=-2C/(D+1)+o(1)$,
\be\label{lemma1}
\lim_{V\to\infty}\frac{1}{V}\inf_{\{g_\k\}}E_{m_0=m_V}=-\frac{\mu^2}{2v(0)},
\ee
the energy density in the ground state of the full diagonal model.

\emph{Proof.} In terms of $C$, $D$ and $x_0$
\be\label{mV}
m_V=-\frac{1}{2}+(D+1-Cx_0)\frac{\sqrt{1-x_0^2}}{1+x_0^2/2}\ .
\ee
Substituting the assumed value of $x_0$ and $\sqrt{1-x_0^2}=1/(1+2h_0)$,
\be
\langle N_0\rangle=h_0+(1+2h_0)m_V=\frac{V}{v(0)}(\mu-\nu_1)+o(V).
\ee
Adding to this the number of uncondensed particles, $\sum_{\k\neq 0}h_\k$,
\be
\langle N\rangle=\frac{\mu V}{v(0)}-\frac{1}{v(0)}\sum_{\k\neq 0}v(\k)h_\k+o(V).
\ee
Suppose now that $g_\k^\mu$ is the energy-minimizing set. Then
\bea\label{dE}
\lefteqn{
\frac{\d E_{m_0=m_V}(\{g_\k^\mu\})}{\d\mu}=-\langle N\rangle}\nonumber\\
&&=-\frac{\mu V}{v(0)}+\frac{1}{v(0)}\sum_{\k\neq 0}v(\k)h_\k^\mu+o(V)
\eea
where the $o(V)$ correction can be negative. On the other hand, at $\mu=0$ we have $g_\k^0=0$ for all $\k$, and $E_{m_0=m_V}(\{g_\k^0\})=0$ (because $\varphi\geq 0$). Integrating Eq.~(\ref{dE}),
\be\label{Egmu}
E_{m_0=m_V}(\{g_\k^\mu\})=-\frac{\mu^2 V}{2v(0)}+\frac{1}{v(0)}\sum_{\k\neq 0}v(\k)\int_0^\mu h_\k^{\mu'}\d\mu'+o(V)
\ee
Because this is the energy minimum,
$$E_{m_0=m_V}(\{g_\k^\mu\})\leq E_{\rm MF}.$$
The second term in the right member of Eq.~(\ref{Egmu}) is nonnegative and, thus, of $o(V)$  [implying also $g_\k=o(1)$], otherwise the above inequality could not hold. This, however, proves Eq.~(\ref{lemma1}).

An energy density smaller than $-\mu^2/2v(0)$ can only be obtained if the energy-minimizing $g_\k$ for $\k\neq 0$ are such that $|2C/(D+1)|>1$ and remains separated from 1 as $V$ tends to infinity. In what follows we shall investigate this possibility.

\emph{Lemma 2.} Suppose that $|2C/(D+1)|>1$ and remains separated from 1 as $V$ tends to infinity. Assume also $\mu-\nu_1+|\nu_2|> 0$. Then the energy-minimizing $\chi_0$ and also $m_V$ are of order $\sqrt{V}$. Furthermore,
\be\label{enVB}
\inf_{g_0}E_{m_0=m_V}=-\frac{V}{3v(0)}(\mu-\nu_1+|\nu_2|)^2+J_{\neg 0}+o(V)
\ee
where $J_{\neg 0}$ is $w_0$ at $g_0=0$,
\bea\label{Jneg0}
J_{\neg 0}=\sum_{\k\neq 0}(\veps(\k)-\mu)h_\k+\frac{v(0)}{2V}\left(\sum_{\k\neq 0}h_{\k}\right)^2 \nonumber\\
+\frac{1}{2V}\sum_{\k,\k'\neq 0}v(\k-\k')(\chi_\k\chi_{\k'}+h_\k h_{\k'})\ .
\eea

\emph{Proof.} Under the condition of the lemma and with the choice $C<0$, $x_0$ is necessarily in $o(1)$ distance to 1. To compute its asymptotic form, in $G_V$ we insert $x_0=1$ everywhere except for $1-x_0^2$ and obtain to leading order
\be
G_V(x_0)=\frac{27}{16}-(D+1+|C|)(2|C|-D-1)(1-x_0^2)^2.
\ee
Solving $G_V(x_0)=0$ yields
\be
1-x_0^2=\frac{3\sqrt{3}}{4\sqrt{(D+1+|C|)(2|C|-D-1)}}\ .
\ee
This shows that $x_0=1-O(V^{-1})$,
\be
\chi_0=\left[\frac{(D+1+|C|)(2|C|-D-1)}{27}\right]^{\frac{1}{4}}\propto\sqrt{V}
\ee
and with $\chi_0=x_0/(2\sqrt{1-x_0^2})$
\be
m_V=-\frac{1}{2}+\frac{D+1+|C|}{3\chi_0}\propto\sqrt{V}\ .
\ee
From here the form (\ref{enVB}) of the energy is obtained by substitution into Eq.~(\ref{Eg0}).

To complete the proof of the theorem it remains to show that (\ref{enVB}) can be smaller than $E_{\rm MF}$ and the difference can be of the order of the volume. Let
\be\label{nu0}
\nu_1=\nu_0+v(0)\rho',\quad\nu_0=\frac{1}{V}\sum_{\k\neq 0}v(\k)h_\k,\quad\rho'=\frac{1}{V}\sum_{\k\neq 0}h_\k
\ee
and
\bea\label{j}
j_{\neg 0}=\frac{1}{V}J_{\neg 0}+\mu\rho'
=\frac{1}{V}\sum_{\k}\veps(\k)h_\k+\frac{v(0)}{2}(\rho')^2 \nonumber\\
+\frac{1}{2V^2}\sum_{\k,\k'\neq 0}v(\k-\k')(\chi_\k\chi_{\k'}+h_\k h_{\k'})\ .
\eea
Now
\bea\label{diff}
\frac{1}{V}\inf_{g_0}E_{m_0=m_V}+\frac{\mu^2}{2v(0)}+o(1)\nonumber\\
=\frac{1}{6v(0)}\{\mu^2-2[v(0)\rho'+2(|\nu_2|-\nu_0)]\mu\nonumber\\
+6v(0)j_{\neg 0}-2(|\nu_2|-\nu_1)^2\}
\eea
is negative if
\be
\Delta=[v(0)\rho'+2(|\nu_2|-\nu_0)]^2+2(|\nu_2|-\nu_1)^2-6v(0)j_{\neg 0}>0
\ee
and
\be\label{muineq}
v(0)\rho'+2(|\nu_2|-\nu_0)-\sqrt{\Delta}<\mu<v(0)\rho'+2(|\nu_2|-\nu_0)+\sqrt{\Delta}.
\ee
The conditions $e_0<0$ (or $m_V>0$) and $2|C|/(D+1)>1$ of the validity of (\ref{enVB}) imply
\bea\label{muineq2}
\lefteqn{\nu_1-|\nu_2|=}\nonumber\\
&&v(0)\rho'+\nu_0-|\nu_2|<\mu<v(0)\rho'+2|\nu_2|+\nu_0\nonumber\\
&&=\nu_1+2|\nu_2|\ .
\eea
The intervals (\ref{muineq}) and (\ref{muineq2}) for $\mu$ clearly overlap if $|\nu_2|>\nu_0$ which can be realized if all $\chi_\k$ have the same (negative) sign because $h_\k<|\chi_\k|$.  We therefore do not consider the otherwise interesting possibility that $\chi_\k$ can vary in sign. On the other hand, for interactions  that are bounded at the origin $\chi_\k\leq 0$ for all $\k\neq 0$ implies that $\Delta$ becomes negative for large $\mu$. Indeed, with Eqs.~(\ref{nu0}) and (\ref{j}) the discriminant can be rewritten in a more convenient form,
\bea
\Delta=6(|\nu_2|-\nu_0)^2-\frac{6v(0)}{V}\sum_{\k}\veps(\k)h_\k \nonumber\\
-\frac{3v(0)}{V^2}\sum_{\k,\k'\neq 0}v(\k-\k')(\chi_\k\chi_{\k'}+h_\k h_{\k'})
\eea
Using the Schwarz inequalities
\bea\label{Schwarz}
\frac{v(0)}{V^2}\sum_{\k,\k'\neq 0}v(\k-\k')\chi_\k\chi_{\k'}\geq\nu_2^2\nonumber\\
\frac{v(0)}{V^2}\sum_{\k,\k'\neq 0}v(\k-\k')h_\k h_{\k'}\geq\nu_0^2
\eea
one obtains
\be
\Delta\leq 3(|\nu_2|-\nu_0)^2-6\nu_0|\nu_2|-\frac{6v(0)}{V}\sum_{\k}\veps(\k)h_\k.
\ee
Now
\bea\label{delta-nu}
0\leq|\nu_2|-\nu_0=\frac{1}{V}\sum_{\k\neq 0}v(\k)(|\chi_\k|-h_\k)\nonumber\\
<\frac{1}{2V}\sum_{\k\neq 0}v(\k)\asymp\frac{1}{2}\varphi(0)
\eea
because $h_\k\leq |\chi_\k|<h_\k+\frac{1}{2}$, and therefore
\be
\Delta<\frac{3}{4}\varphi(0)^2-6\nu_0|\nu_2|-\frac{6v(0)}{V}\sum_{\k}\veps(\k)h_\k.
\ee
It is then clear that the positivity of $\Delta$ places an upper bound on $|\chi_\k|$ and, via (\ref{muineq}) or (\ref{muineq2}), on $\mu$.

We now show that under some additional condition on the interaction, for any small enough $\mu$ the difference (\ref{diff}) can be made negative. Because $\varphi\geq 0$ and is integrable, $v(\k)$ is continuous and $|v(\k)|\leq v(0)$. Choose $0<\gamma<1$ and $k_\gamma>0$ such that $v(0)\geq v(\k)\geq\gamma v(0)$ for $k\leq k_\gamma$. Choose moreover $\chi_\k\leq 0$ and
\be\label{ansatz}
h_\k=\left\{\begin{array}{ll}
                  h &\mbox{if $0<k\leq k_\gamma$}\\
                  0    & \mbox{if $k>k_\gamma$}
                  \end{array}\right.
\ee
Let
\be
B_\gamma=\frac{1}{V}\sum_{k\leq k_\gamma}1\approx\frac{k_\gamma^3}{6\pi^2}.
\ee
Then with $\chi=\sqrt{h(h+1)}$,
\be
\frac{1}{V}\sum_{k}\veps(\k)h_\k\approx\frac{3}{5}hB_\gamma\veps(\k_\gamma)=\frac{\sigma h k_\gamma^5}{10\pi^2}
\ee
($\sigma=\hbar^2/2m$) and
\be
\gamma B_\gamma v(0)\leq \frac{1}{V}\sum_{k\leq k_\gamma}v(\k)\leq B_\gamma v(0)
\ee
we obtain
\be\label{deltalower}
\frac{\Delta}{3hB_\gamma^2v(0)^2}\geq 2\gamma^2-1-\frac{(6\pi)^2\sigma}{5v(0)k_\gamma}-4\gamma^2(\chi-h)-2h.
\ee
The lower bound is positive for small enough $h$ if
\be
5(2\gamma^2-1)v(0)k_\gamma>(6\pi)^2\sigma.
\ee
This shows that $\gamma>1/\sqrt{2}$ must be. For example, with the choice $\gamma=0.9$ and $k_\gamma=(4\pi)^2\sigma/v(0)$ the condition on the interaction reads
\be\label{cond-v}
v(\k)\geq 0.9v(0)\quad{\rm for}\quad k\leq\frac{(4\pi)^2\sigma}{v(0)}\ .
\ee
The positivity of the r.h.s. of (\ref{deltalower}) imposes $\chi<1/\sqrt{12}$, but no lower bound on $\chi$. Therefore the (positive) lower edge of the interval (\ref{muineq}) can be arbitrarily close to zero. Similar holds to the lower edge of the interval (\ref{muineq2}) if it turns out to be positive. Thus, under condition (\ref{cond-v}) the difference (\ref{diff}) is indeed negative for any small positive $\mu$. Although $\varphi(0)<\infty$ for the Gaussian example (\ref{example}), Eq.~(\ref{cond-v}) is compatible with the non-integrability of $v$ and, hence, with $\varphi(0)=\infty$.

\subsection{Gap to excitations in the VB scheme}

Adding $n$ quasiparticles of wave vector $\k=0$ to the QPGS needs an energy
\be\label{gapg0}
E_{m_0=m_V\pm n}-E_{m_0=m_V}=w_{00}n^2
\ee
while the gap to $\k\neq 0$ excitations is
\bea\label{gapgk}
\lefteqn{E_{m_0=m_V,m_\k=1}-E_{m_0=m_V}}\nonumber\\
&&=e_\k+2w_{0\k}m_V=e_\k-\frac{w_{0\k}}{w_{00}}e_0+w_{0\k}.
\eea
Equations (\ref{gap00})-(\ref{gap0k}) are special cases of Eqs.~(\ref{gapg0})-(\ref{gapgk}), obtained with all $g_\k=0$. If the QPGS is characterized by Lemma 2, the gap to $\k=0$ excitations is
\bea
\lefteqn{
w_{00}=\frac{v(0)}{2V}(1+6\chi_0^2)\approx \frac{3v(0)}{V}\chi_0^2}\nonumber\\
&&=\sqrt{\frac{1}{3}(\mu-\nu_1+|\nu_2|)(-\mu+\nu_1+2|\nu_2|)}.
\eea
To compute (\ref{gapgk}), $E_{m_0=m_V}$ has to be minimized w.r.t. $\chi_\k$. The analogous procedure in the NStJAW scheme will be discussed in detail in the next section. Here we present only the limit of (\ref{gapgk}) as $\k\to 0$ \cite{comment2}. It can be obtained from
\be
\lim_{\k\to 0}x_\k=-\frac{2(\mu-\nu_1)-|\nu_2|}{\mu-\nu_1+4|\nu_2|}
\ee
valid for the minimizing $x_\k$, cf. Eq.~(\ref{xk}). The result is
\bea
\lefteqn{
\lim_{\k\to 0}\left(E_{m_0=m_V,m_\k=1}-E_{m_0=m_V}\right)}\nonumber\\
&&=\sqrt{\frac{1}{3}(\mu-\nu_1+|\nu_2|)(-\mu+\nu_1+5|\nu_2|)}.
\eea
This is somewhat higher than $w_{00}$, the gap to $\k=0$ excitations. The discontinuity of the gap comes from the discontinuity of $x_\k$ at $\k=0$. Indeed, from the strict inequalities (\ref{muineq2}) one obtains
$$\lim_{\k\to 0}x_\k<x_0=1-O(V^{-1}).$$

\newsec{The Nozi\`eres-Saint James-Araki-Woods scheme}
\subsection{Quasiparticle ground state}

Comparing Eq.~(\ref{enVB}) with the lower bound (\ref{lb}) one observes a prefactor $\frac{1}{3}$ in the former instead of $\frac{1}{2}$ in the latter. The source of the difference is $D^2(N_0)$ which was replaced by 0 when (\ref{lb}) was derived but which is
\be
D^2(N_0)\approx\frac{1}{2}\langle N_0\rangle^2\approx 2\chi_0^2 m_V^2\approx\frac{2V^2}{9v(0)^2}(\mu-\nu_1+|\nu_2|)^2
\ee
in case when (\ref{enVB}) is valid. In the VB scheme the fluctuation of the number of condensed particles is anomalously large, manifesting itself in an excess energy density. In 1982 Nozi\`eres and Saint James~\cite{NStJ} studied a modification of the VB scheme in which they dropped the Bogoliubov transformation in the zero mode and applied instead Bogoliubov's $c$-number substitution for $a_0$. In the present work this corresponds to setting $g_0=0$ and applying a shift (\ref{shift}) only in the zero mode: $\tilda_0=a_0-z=b_0$, $\tilda_\k=a_\k$ for $\k\neq 0$. Because then
\be
D^2(N_0)=\langle N_0\rangle=|z|^2,
\ee
by minimizing the energy w.r.t. $z$, up to a correction of order 1 the lower bound (\ref{lb}) could be attained. As Eq.~(\ref{expN0}) suggests, the minimizer was
\be\label{minimz}
|z|^2=\frac{V}{v(0)}(\mu-\nu_1+|\nu_2|),
\ee
see below. Thus, remaining within the family of pairing product states, only an improvement of order 1 can be gained on the NStJ result. It is nonetheless instructive to see the origin and some consequence of this improvement. Before arriving there, first we will have a closer look at the NStJ setting. Nozi\`eres and Saint James considered only the vacuum state $\Phi_0$. Here we take into account all states of the basis (\ref{basis}) and prove the following.

\emph{Theorem 2.} Let $\varphi\geq 0$ and $v\geq 0$. Consider the basis (\ref{basis}) with $g_0=0$ and $z_\k=z\delta_{\k,0}$. Under the condition of uniqueness, the QPGS is a quasiparticle vacuum state with an energy density smaller than $-\mu^2/2v(0)$ for all $\mu>0$. If $\varphi(0)=\infty$, for $\mu$ large the full particle density is $\rho=[\mu+o(\mu)]/v(0)$ where the $o(\mu)$ term tends to infinity with $\mu$. If $\varphi(0)<\infty$, for $\mu$ large $\rho=[\mu+\varphi(0)/2]/v(0)+o(\mu^0)$. In all cases the uncondensed density  tends to infinity with $\mu$ but slower than $\mu$, e.g. $\rho'\sim\mu^{1/3}$ if the decay of $v$ is fast enough.

\emph{Remarks.} (i) In the ground state of a system of classical particles whose interaction satisfies $v\geq 0$, $v(0)>0$ and $v(\k)=0$ for $k>k_0$, the exact relation $\rho=[\mu+\varphi(0)/2]/v(0)$ holds above a threshold density $\sim k_0^3$, see Ref.~\cite{Su3}.

\noindent
(ii) The role of $\varphi\geq 0$ is to guarantee that the QPGS \emph{is} a vacuum state -- just the opposite of its role in the VB scheme.

\emph{Proof.} The argument given in Proposition 2 remaining valid, uniqueness implies that in the QPGS only $m_0$ can be nonzero. Let $\Phi_{m_0}$ denote such an element of the basis and $\langle B\rangle$ the expectation value of the operator $B$ in this state. For $\k\neq 0$ from (\ref{avfirst})-(\ref{avlast}) we find $\langle N_\k\rangle=h_\k$, $\langle A_\k\rangle=\chi_\k$ and $D^2(N_\k)=\chi_\k^2$. Because $m_\k=0$, $\langle\Hnot0\rangle=J_{\neg 0}$, cf. Eq.~(\ref{Jneg0}). For $\k=0$ now we have
\be
\langle N_0\rangle=|z|^2+m_0,\qquad\langle A_0\rangle=z^2
\ee
and
\be
D^2(N_0)=|z|^2(1+2m_0)=\langle N_0\rangle+m_0(2|z|^2-1).
\ee
Choose $z$ real and $\nu_2\leq 0$. Insert $\langle N_0\rangle-\langle A_0\rangle$  and $D^2(N_0)$ from above into (\ref{avH0}) to find
\bea\label{enNStJ}
\langle H\rangle=J_{\neg 0}-(\mu-\nu_1+|\nu_2|)\langle N_0\rangle+\frac{v(0)}{2V}\langle N_0\rangle^2
\nonumber\\
+m_0\left(|\nu_2|+ \frac{v(0)}{V}z^2-\frac{v(0)}{2V}\right).
\eea
If $|\nu_2|$ is of order 1, the coefficient of $m_0$ is positive for any $z$, and the minimum of $\langle H\rangle$ is obtained with $m_0=0$ and, provided that $\mu-\nu_1+|\nu_2|\geq 0$, with $\langle N_0\rangle=z^2$ given by (\ref{minimz}). Its value is
\bea\label{minvacuum}
E_0=\min_z(\Phi_0,H\Phi_0)=-\frac{V}{2v(0)}(\mu-\nu_1+|\nu_2|)^2+J_{\neg 0}\nonumber\\
=-\frac{V}{2v(0)}(\mu-\nu_0-\nu_2)^2-(\nu_0+\nu_2)\rho'V+\sum\veps(\k)h_\k\nonumber\\
+\frac{1}{2V}\sum_{\k,\k'\neq 0}v(\k-\k')(\chi_\k\chi_{\k'}+h_\k h_{\k'}),\phantom{aaaaaaaaaa.}
\eea
see Eqs.~(\ref{nu0}), (\ref{j}). Note that
\be\label{rho}
\rho=\langle N\rangle/V=\rho_0+\rho'=\frac{\mu+|\nu_2|-\nu_0}{v(0)},
\ee
thus, the first term of the last member of (\ref{minvacuum}) is $-v(0)\rho^2V/2$. The minimum of $E_0$ over $\{g_\k\}_{\k\neq 0}$ is to be compared with the lowest energy one can obtain if the coefficient of $m_0$ in (\ref{enNStJ}) is negative.

We show that if the coefficient of $m_0$ in (\ref{enNStJ}) is negative then the energy is at the best by order 1 below $-\mu^2V/2v(0)$. In this case $z^2<\frac{1}{2}$ and $|\nu_2|<\frac{v(0)}{2V}$. (The latter may occur either because $\chi_\k$ are small or because they oscillate in sign, in which case $\nu_1$ may still be of order 1.) Because of the superstability of the interaction, in the QPGS $m_0$ can be at most of order $V$. Therefore, the contribution to $\langle H\rangle$ of the term proportional to $m_0$ is at most of order 1. For the same reason $|\nu_2|\langle N_0\rangle=O(1)$. As for the rest, there are two cases to be distinguished. (i) $\mu-\nu_1\leq 0$. Then the minimum of $\langle H\rangle$ is, apart from a correction of order 1, that of $J_{\neg 0}$ which cannot be smaller than $-\mu^2V/2v(0)$, cf. Eq.~(\ref{w0bis}). (ii) $\mu-\nu_1>0$. Minimizing the right member of Eq.~(\ref{enNStJ}) with respect to $\langle N_0\rangle$,
\be
\langle H\rangle +O(1)\geq J_{\neg 0}-\frac{V}{2v(0)}(\mu-\nu_1)^2.
\ee
Substituting $J_{\neg 0}$ from (\ref{Jneg0}), $\nu_1=\nu_0+v(0)\rho'$, and applying the Schwarz inequalities (\ref{Schwarz}), after simplification we find
\bea
\langle H\rangle +O(1)&\geq& \sum\veps(\k)h_\k+\frac{V}{v(0)}(\mu-v(0)\rho')\nu_0-\frac{\mu^2V}{2v(0)}
\nonumber\\
&\geq& -\frac{\mu^2V}{2v(0)}
\eea
because $\mu-v(0)\rho'>\nu_0\geq 0$.

On the other hand, by choosing $\chi_\k\leq 0$ and $|\nu_2|$ small but non-vanishing as $V$ tends to infinity, for any $\mu>0$ the energy (\ref{minvacuum}) can be made smaller than $-\mu^2V/2v(0)$ by an amount of order $V$. Indeed, in this case $-\nu_0-\nu_2\geq 0$, cf. Eq.~(\ref{delta-nu}), so an improvement comes from the first term of (\ref{minvacuum}); the last three terms are positive, independent of $\mu$ and can be made arbitrarily small by decreasing $|\chi_\k|$. Consider, for example, the choice (\ref{ansatz}). Because $h<\chi^2$, in the limit when $\chi\to 0$ we can drop terms of order $\chi^2$ or smaller in (\ref{minvacuum}). Therefore
\be\label{small-chi}
\frac{E_0}{V}=-\frac{\mu^2}{2v(0)}-\frac{\mu|\nu_2|}{v(0)}+O(\chi^2)
\leq -\frac{\mu^2}{2v(0)}-\gamma B_\gamma\chi\mu+O(\chi^2).
\ee

While (\ref{small-chi}) is the relevant formula for the energy density if $\mu$ is small, the situation is quite different in the other asymptotic region, when $\mu$ tends to infinity. It is important to realize that $\chi_\k$ minimizing the energy do not remain small when $\mu$ increases, as the argument leading to (\ref{small-chi}) would suggest it. On the contrary, they tend to infinity with $\mu$, although more slowly. This will become obvious when we minimize (\ref{minvacuum}) in the variables $\chi_\k$. From Eq.~(\ref{minvacuum}),
\be\label{partial-E0}
\frac{\partial E_0}{\partial\chi_\k}=2x_\k[v(\k)\rho_0+\nu_{0\k}-\nu_0-\nu_2+\veps(\k)]
+2[v(\k)\rho_0+\nu_{2\k}]
\ee
where
\be\label{rho-0}
\rho_0=\frac{\langle N_0\rangle}{V}=\frac{\mu-\nu_1-\nu_2}{v(0)}
\ee
and
\be
\nu_{0\k}=\frac{1}{V}\sum_{\k'\neq 0}v(\k-\k')h_{\k'},\quad
\nu_{2\k}=\frac{1}{V}\sum_{\k'\neq 0}v(\k-\k')\chi_{\k'}
\ee
($\nu_{00}=\nu_0$, $\nu_{20}=\nu_{2}$, $\nu_{i,-\k}=\nu_{i\k}$.) Then $\partial E_0/\partial\chi_\k=0$ yields
\be\label{opt-xk}
x_\k=-\frac{v(\k)\rho_0+\nu_{2\k}}{v(\k)\rho_0+\nu_{0\k}-\nu_0-\nu_2+\veps(\k)}.
\ee
This is still an implicit relation because $\rho_0,\nu_0,\nu_{0\k},\nu_2,\nu_{2\k}$ depend on all $\chi_{\k'}$. Let $\mu$ tend to infinity and assume that $|\chi_{\k'}|$ do not increase. Because $\rho_0$ tends to infinity, $x_\k\to~-1$, i.e., $\chi_\k\to-\infty$, contradicting the assumption. So $|\chi_{\k'}|$ do increase and therefore the uncondensed density $\rho'$ and all the $\nu$ diverge with a diverging $\mu$. To allow $x_\k$ to tend to $-1$, $|\nu_{i\k}|/\rho_0$ must tend to zero. Moreover, because $\nu_{0\k}$ and $\rho'$ are of the same order, $\rho'/\rho_0$ tends to zero as $\mu$ tends to infinity. Comparison with (\ref{rho-0}) then shows that $\rho_0$ is of the order of $\mu$ and $\rho'=o(\mu)$. The result for $\rho$ follows from Eqs.~(\ref{rho}) and (\ref{delta-nu}) with the remark that
\be\label{expansion}
|\chi_\k|-h_\k=\frac{1}{2}-\frac{1}{8h_\k}+O\left(\frac{1}{h_\k^2}\right)
\ee
as $h_\k\to\infty$.

More precise asymptotics in $\mu$ can be obtained if we choose $h_\k=\lambda h(\k)$ and $\chi_\k\leq 0$ where $h(\k)>0$ satisfies $\int\veps(\k)h(\k)\d\k<\infty$ and $\int~v(\k)/h(\k)^2\d\k<\infty$. For such an $h(\k)$ to exist $v$ must decay fast enough; for example, a Gaussian interaction is suitable. Now $\lambda$ is a parameter to minimize $E_0$. As $\mu\to\infty$, the minimizing $\lambda$ tends to infinity as well. The terms of $E_0/V$ determining the asymptotic form of $\lambda(\mu)$ are
\[
\frac{\mu\int[v(\k)/h(\k)]\d\k}{(4\pi^3)v(0)\lambda}+\frac{\lambda^2}{(2\pi)^6}
\int\int v(\k-\k')h(\k)h(\k')\d\k\d\k'
\]
where the first term comes from the expansion (\ref{expansion}). The minimum is attained at
\be
\lambda=\pi\left[\frac{\mu\int[v(\k)/h(\k)]\d\k}{2v(0)\int\int v(\k-\k')h(\k)h(\k')\d\k\d\k'}\right]^\frac{1}{3}.
\ee
We find therefore $\rho'=(\lambda/8\pi^3)\int h(\k)\d\k\sim\mu^\frac{1}{3}$. This finishes the proof of the theorem.

\subsection{Quasiparticle Hamiltonian and gap to excitations in the NStJAW scheme}

With the help of the basis (\ref{basis}) one can again define a quasiparticle Hamiltonian. The contribution of $H_{\neg 0}$ to $H_{\rm QP}$ is the same as in the VB scheme, the difference comes from a different contribution of $H_0$. Compared with (\ref{Hg}) the changes are as follows. First set $h_0=\chi_0=0$ at every occurrence. This yields
\be
w_0=J_{\neg 0},\quad e_0=-\mu+\nu_1,\quad w_{00}=\frac{v(0)}{2V},
\ee
\be
e_\k=(1+2h_\k)(\veps(\k)-\mu+\nu_{1\k})+2\chi_\k\nu_{2\k}
\ee
and
\be
w_{0\k}=\frac{1}{2V}(1+2h_\k)[v(0)+v(\k)]\quad(\k\neq 0)
\ee
where
\be
\nu_{1\k}=v(0)\rho'+\nu_{0\k}
\ee
($\nu_{10}=\nu_1$, $\nu_{1,-\k}=\nu_{1\k}$). Second, replace $w_0$, $e_{\k\neq 0}$ and $e_0$ respectively by
\be\label{VBtilde}
\widetilde{w}_0=w_0-(\mu-\nu_1)|z|^2+({\rm Re\,}z^2)\nu_2+\frac{v(0)}{2V}|z|^4,
\ee
\be
\widetilde{e}_\k=e_\k+\frac{|z|^2}{V}(1+2h_\k)[v(0)+v(\k)]+\frac{2\chi_\k}{V}v(\k){\rm Re\,}z^2
\ee
and
\be
\widetilde{e}_0=e_0+\frac{2v(0)}{V}|z|^2.
\ee
The other coefficients are unchanged. Because the ground state of $H_{\rm QP}$ is now $\Phi_0$, adding a quasiparticle of wave vector $\k$ to the ground state costs an energy $\widetilde{e}_\k$. Inserting $z$ real from Eq.~(\ref{minimz}), using Eq.~(\ref{rho-0}) and $\nu_{1\k}-\nu_1=\nu_{0\k}-\nu_0$,
\be
\widetilde{e}_\k=(1+2h_\k)[v(\k)\rho_0(1+x_\k)+\nu_{0\k}-\nu_0
+x_\k\nu_{2\k}-\nu_2+\veps(\k)]
\ee
and
\be
\widetilde{e}_0=\mu-\nu_1-2\nu_2=v(0)\rho_0+|\nu_2|.
\ee
From (\ref{opt-xk})
\be\label{lim-xk}
\lim_{\k\to 0}x_\k=-\frac{v(0)\rho_0+\nu_{2}}{v(0)\rho_0-\nu_2}=-\frac{\mu-\nu_1}{\mu-\nu_1-2\nu_2}
\ee
see the remark \cite{comment2}. Substituting this into $\widetilde{e}_\k$ and using $1+2h_\k=1/\sqrt{1-x_\k^2}$ we find
\be\label{lim-ek}
\lim_{\k\to 0}\widetilde{e}_\k=2\sqrt{v(0)\rho_0|\nu_2|}>0,
\ee
the same as the gap value in Refs. \cite{GA} and \cite{NStJ}. As expected, $\lim_{\k\to 0}\widetilde{e}_\k\neq\widetilde{e}_0$ because $\lim_{\k\to 0}x_\k\neq x_0=0$. In effect,
\be
0<\lim_{\k\to 0}\widetilde{e}_\k<\widetilde{e}_0.
\ee
The discontinuity of the gap at $\k=0$ will be removed in the next section.

In terms of the numerator and the denumerator of $x_\k$, $\widetilde{e}_\k$ has a nice form. Let
\be
\alpha_\k=v(\k)\rho_0+\nu_{2\k},\quad\beta_\k=v(\k)\rho_0+\nu_{0\k}-\nu_0-\nu_2+\veps(\k).
\ee
Then
\be\label{xk-ek}
x_\k=-\alpha_\k/\beta_\k,\quad\widetilde{e}_\k=\sqrt{(\beta_\k+\alpha_\k)(\beta_\k-\alpha_\k)},
\ee
Choose now $\k\neq 0$ such that
\be\label{interval}
v(0)\rho'\ll\veps(\k)\ll v(0)\rho_0.
\ee
Keeping only the largest term,
\be
\widetilde{e}_\k\approx\sqrt{2v(\k)\rho_0\veps(\k)}.
\ee
This is Bogoliubov's dispersion relation that was reproduced already by Girardeau and Arnowitt \cite{GA} and is obtained here within the NStJAW scheme. With energy-minimizing $\chi_\k$ the interval (\ref{interval}) does not extend down to zero because the optimal $\rho'$ is separated from zero. Hence, the linear dispersion exists only as an extrapolation near $\k=0$. This, of course, is consistent with Eq.~(\ref{lim-ek}). As $\mu$ diverges, the interval $[v(0)\rho',v(0)\rho_0]$ widens and its lower bound shifts upwards e.g. as $\mu^{1/3}$, cf. Theorem~2. As $\mu$ tends to zero, the interval gets closer to zero. Because of the relation $v(0)\rho_0=\mu-\nu_1+|\nu_2|$, the decrease of $\mu$ is consistent also with $\rho_0$ fixed and $v(0)$ and all $h_\k$, $|\chi_\k|$ and, therefore, $\rho'$ decreasing. It is in this sense that Bogoliubov's result for a weakly interacting Bose gas with an almost complete BEC can be recovered. When we take the limit $v(0)\to 0$, Eq.~(\ref{opt-xk}) shows that $x_\k\to 0$ (and, thus, $\langle N_\k\rangle\to 0$) for any \emph{fixed} nonzero $\k$. However, if $\k$ decreases with $v(0)$ so that $\veps(\k)$ remains in the interval (\ref{interval}), then $x_\k$ does not tend to zero; actually $\lim|x_\k|>1/2$ in this joint limit. Note that the joint limit can be realized only in infinite volume, when $\k$ can vary continuously, and the result does not contradict $\rho'\to 0$. In finite volumes, for $v(0)\rho_0<h^2/2mL^2$ there is no $\veps(\k)$ in the interval (\ref{interval}), and $x_\k$ eventually tends to zero with $v(0)$ for every nonzero $\k$.

\subsection{$z_0\neq 0$ and $g_0\neq 0$}

Here we discuss a slight improvement of the NStJAW scheme. Comparison of Eq.~(\ref{minvacuum}) with the lower bound (\ref{lb}) shows that by letting $g_0$ be nonzero and by minimizing the energy with respect to it, only a gain of the order of unity can be attained. Minimization in $g_0$ is, nonetheless, interesting because it makes the excitation energy continuous at $\k=0$.

Choosing $g_0$ nonzero affects only $\langle H_0\rangle$ which can be computed from Eq.~(\ref{avH0}). The relevant values to be substituted here are
\be\label{0-entries1}
\langle N_0\rangle=z^2+\langle\widetilde{N}_0\rangle,\qquad\langle A_0\rangle=z^2+\langle\widetilde{A}_0\rangle
\ee
and
\be\label{0-entries2}
D^2(N_0)=z^2[1+2\langle\widetilde{N}_0\rangle]+2z^2\langle\widetilde{A}_0\rangle
\ee
where $z$ is real and
\bea
\langle\widetilde{N}_0\rangle&=&\langle\tilda_0^*\tilda_0\rangle=h_0+(1+2h_0)m_0,\nonumber\\
\langle\widetilde{A}_0\rangle&=&\langle\tilda_0^2\rangle=\chi_0+2\chi_0 m_0.
\eea
These are averages in a state (\ref{basis}) with $z_\k=m_\k=0$ for $\k\neq 0$ and $z_0=z$; $h_0$ and $\chi_0$ are related to $g_0$ via Eq.~(\ref{h-chi}). Substituting (\ref{0-entries1}) and (\ref{0-entries2}) into Eq.~(\ref{avH0}) we find
\bea
\langle H_0\rangle=-\left(\mu-\nu_1-\nu_2-\frac{v(0)}{V}[\langle \widetilde{N}_0\rangle+\langle \widetilde{A}_0\rangle]\right)\,\langle N_0\rangle\nonumber\\
+\frac{v(0)}{2V}\langle N_0\rangle^2-\nu_2\,[\langle \widetilde{N}_0\rangle-\langle \widetilde{A}_0\rangle]\nonumber\\
-\frac{v(0)}{2V}\langle \widetilde{N}_0\rangle[1+2\langle \widetilde{N}_0\rangle+2\langle \widetilde{A}_0\rangle]\nonumber\\
\phantom{a}
\eea

The minimum of $\langle H_0\rangle$ as a function of $\langle N_0\rangle$ reads
\bea\label{H0-with-g0-z}
\langle H_0\rangle=-\frac{v(0)}{2V}\langle N_0\rangle^2
-\nu_2[\langle\widetilde{N}_0\rangle-\langle\widetilde{A}_0\rangle]\nonumber\\
-\frac{v(0)}{2V}\langle\widetilde{N}_0\rangle
[1+2\langle\widetilde{N}_0\rangle+2\langle\widetilde{A}_0\rangle]
\eea
obtained with
\be\label{N0-beyond}
\langle\label{ro0-beyond} N_0\rangle=V\rho_0=\frac{V}{v(0)}(\mu-\nu_1-\nu_2)-\langle\widetilde{N}_0\rangle-\langle\widetilde{A}_0\rangle.
\ee
The $m_0$-dependent part of $\langle H_0\rangle$ in (\ref{H0-with-g0-z}) reads
\bea\label{m0-part}
(1+2h_0)\left[|\nu_2|(1-x_0)-\frac{v(0)}{2V}(1+4h_0)(1+x_0)\right]m_0\nonumber\\
-\frac{v(0)}{V}(1+2h_0)^2(1+x_0)m_0^2.\nonumber\\
\phantom{A}
\eea
The difference $E_0-\langle H_0\rangle$, where $E_0$ and $\langle H_0\rangle$ are taken from Eqs.~(\ref{minvacuum}) and (\ref{H0-with-g0-z}), respectively, cannot be larger than of order $V^0$. This, however, means that $\langle\widetilde{N}_0\rangle$ and $h_0$ and $m_0$ together with it are of order $V^0$, implying that the expression (\ref{m0-part}) is nonnegative for $V$ large enough and its minimum is attained at $m_0=0$. Setting $m_0=0$ we obtain the vacuum expectation value
\bea\label{E0-beyond}
E_0=(\Phi_0,H\Phi_0)=J_{\neg 0}-\frac{V}{2v(0)}(\mu-\nu_1-\nu_2)^2\nonumber\\
+(\mu-\nu_1)(\chi_0+h_0)-2\nu_2 h_0-\frac{v(0)}{V}h_0(1+2h_0+2\chi_0).\nonumber\\
\phantom{a}
\eea
Here the second line contains the correction to (\ref{minvacuum}) which is negative if $\chi_0<0$. For minimization consider
\bea
\frac{\partial E_0}{\partial\chi_0}=
x_0\left[\mu-\nu_1-2\nu_2-\frac{v(0)}{V}(1+4h_0+2\chi_0)\right]\nonumber\\
+\left[\mu-\nu_1-\frac{2v(0)}{V}h_0\right].
\eea
$\partial E_0/\partial\chi_0=0$ yields
\bea\label{x0-beyond}
x_0=-\frac{\mu-\nu_1-\frac{2v(0)}{V}h_0}{\mu-\nu_1-2\nu_2-\frac{v(0)}{V}(1+4h_0+2\chi_0)}\nonumber\\
=-\frac{v(0)\rho_0+\nu_2-\frac{v(0)}{V}(h_0-\chi_0)}{v(0)\rho_0-\nu_2-\frac{v(0)}{V}(1+3h_0+\chi_0)}.
\eea

In computing $\partial E_0/\partial\chi_\k$ for $\k\neq 0$, (\ref{partial-E0}) must be replaced by the derivative of (\ref{E0-beyond}). We find
\be
\frac{\partial E_0}{\partial\chi_\k}=\left.\frac{\partial E_0}{\partial\chi_\k}\right|_{g_0=0}
-2x_\k\frac{v(0)+v(\k)}{V}(\chi_0+h_0)-\frac{4v(\k)}{V}h_0
\ee
where the first term is given by (\ref{partial-E0}). Equating this with zero we obtain
\be\label{xk-beyond}
x_\k=-\frac{v(\k)\rho_0+\nu_{2\k}-\frac{v(\k)}{V}(h_0-\chi_0)}{v(\k)\rho_0+\nu_{0\k}-\nu_0-\nu_2+\veps(\k)
-\frac{v(0)}{V}(h_0+\chi_0)}.
\ee
Note that in Eqs.~(\ref{x0-beyond}) and (\ref{xk-beyond}) $\rho_0$ is substituted from (\ref{ro0-beyond}). It is seen that
\be
\lim_{\k\to 0}x_\k=x_0+O(V^{-1}),
\ee
cf. remark \cite{comment2}.

Consider now the associated quasiparticle theory. $H_{\rm QP}$ is obtained from (\ref{Hg}) by replacing the VB coefficients $w_0$, $e_{\k\neq0}$ and $e_0$ (all defined with $g_0\neq 0$) with the expressions
\bea\label{VBtilde-bis}
\widetilde{w}_0&=&w_0-(\mu-\nu_1-\nu_2)z^2+\frac{v(0)}{2V}z^4+\frac{v(0)}{V}z^2(2h_0+\chi_0)\nonumber\\
\widetilde{e}_\k&=&e_\k+\frac{z^2}{V}(1+2h_\k)[v(0)+v(\k)]+\frac{2\chi_\k}{V}v(\k)z^2\nonumber\\
\widetilde{e}_0&=&e_0+\frac{2v(0)}{V}z^2(1+2h_0+\chi_0)
\eea
where $z$ is real, and keeping the other coefficients unchanged. Now $\langle N_0\rangle=z^2+h_0$, therefore $z^2/V=\rho_0+O(V^{-1})$. Thus, with (\ref{N0-beyond})
\bea\label{tilde-beyond}
\widetilde{e}_\k&=&(1+2h_\k)[v(\k)\rho_0(1+x_\k)+\nu_{0\k}-\nu_0\nonumber\\
&&+x_\k\nu_{2\k}-\nu_2+\veps(\k)]+O(V^{-1})
\eea
while
\be
\widetilde{e}_0=(1+2h_0)[v(0)\rho_0(1+x_0)-\nu_2(1-x_0)]+O(V^{-1}).
\ee
Comparing the above two equations we find
\be
\lim_{\k\to 0}\widetilde{e}_\k=\widetilde{e}_0+O(V^{-1})=2\sqrt{v(0)\rho_0|\nu_2|}+O(V^{-1}),
\ee
the continuity of $\widetilde{e}_\k$ at $\k=0$, up to an error of order $V^{-1}$.

\newsec{Summary}

In this paper we have investigated variational wave functions for the ground state of interacting homogenous Bose systems. We have not invented new trial functions but studied with much care those introduced by Valatin and Butler~\cite{VB} and modified later by Nozi\`eres and Saint James~\cite{NStJ} | though we completed the trial states of the latter into a basis in Fock space. Both sets of trial functions are quasiparticle product states grouped in bases in Fock space, each basis corresponding to a choice of one parameter for every pair $\{\k,-\k\}$. Their difference is in the choice of this parameter for $\k=0$, associated with Bogoliubov's canonical transformation for VB and with a coherent state for NStJ. Each basis consists of a quasiparticle vacuum state and excited states with an arbitrary finite number of quasiparticles. However, it is not \emph{a priori} guarantied that the vacuum state has the lowest energy within a basis, therefore energy minimization must extend to the full basis for a given parameter set. To solve this problem, one has to set precise conditions on the interaction, which in our case are positivity of both the pair interaction and its Fourier transform. Assuming that the overall energy minimum is attained on a single state, we have shown that this state could contain at most quasiparticles of zero momentum. It then turns out that in the VB scheme the lowest energy state is never a vacuum state, as was supposed by Valatin and Butler. For slowly decaying $v(\k)$ and small $\mu$ it is a state with $\sim\sqrt{V}$ quasiparticles, and we could not exclude the possibility that in other cases it is an eigenstate of the number operator $N_0$, namely, the ground state of the corresponding mean field model defined by $v_{\rm MF}(\k)=v(0)\delta_{\k,0}$. In the first case both the mean and the standard deviation of the number of \emph{physical} particles of zero momentum are proportional to the volume. This anomalous fluctuation causes an excess energy density, not present when a coherent state is used in the zero mode. Therefore, the lowest attainable energy density in the NStJAW scheme is always below the minimum obtained in the VB scheme, and is provided by a vacuum state.

Unbiased minimization of the energy leads to a gap in the spectrum of the associated quasiparticle Hamiltonian. We have computed these gaps both at $\k=0$ and in the limit $\k\to 0$. The two values do not coincide in either schemes. In the case of the NStJAW scheme we have shown that the continuity of the gap at $\k=0$ can be restored through an additional energy-minimizing Bogoliubov transformation in the zero mode. We have also discussed how Bogoliubov's linear dispersion relation for a weakly interacting and almost completely condensed Bose gas can be obtained within the NStJAW approach, in spite of the gap in the spectrum.

\section*{Acknowledgment}

This work was supported by OTKA Grant No. T046129, as well as by the program QUDEDIS of the European Science Foundation.


\begin{thebibliography}{99}
\bibitem{comment2}
By $\lim_{\k\to 0}x_\k$ in finite volumes we understand to evaluate the continuous functions $v(\k)$, $v(\k-\k')$, $\veps(\k)$, appearing in $x_\k$, at $\k=0$. Similar for other quantities labelled by $\k$.
\end{thebibliography}
\end{document}